# A picture is worth a thousand words: an empirical study on the influence of content visibility on diffusion processes within a virtual world

Jarosław Jankowski[a,b], Piotr Bródka[b] and Juho Hamari[c]

[a]Faculty of Computer Science and Information Technology, West Pomeranian University of Technology, Szczecin, Poland; [b]Department of Computational Intelligence, Wroclaw University of Technology, Wrocław, Poland; [c]Game Research Lab, School of Information Sciences, University of Tampere, Tampere, Finland

**ABSTRACT**

Studying information diffusion and the spread of goods in the real world and in many digital services can be extremely difficult since information about the information flows is challenging to accurately track. How information spreads has commonly been analysed from the perspective of homophily, social influence, and initial seed selection. However, in virtual worlds and virtual economies, the movements of information and goods can be precisely tracked. Therefore, these environments create laboratories for the accurate study of information diffusion characteristics that have been difficult to study in prior research. In this paper, we study how content visibility as well as sender and receiver characteristics, the relationship between them, and the types of multilayer social network layers affect content absorption and diffusion in virtual world. The results show that prior visibility of distributed content is the strongest predictor of content adoption and its further spread across networks. Among other analysed factors, the mechanics of diffusion, content quality, and content adoption by users' neighbours on the social activity layer had very strong influences on the adoption of new content.

## 1. Introduction

Virtual worlds pose great scientific potential as they can be used as virtual laboratories for observing complex social and economic phenomena (Bainbridge 2007). This is true especially because virtual world technology not only allows real-time synchronous communication and economic activity between users (who are commonly represented usually by graphical avatars), but also allows for the accurate logging of even the smallest economic activities that occur. Because of these qualities of virtual worlds, they hold great potential especially for studying multidimensional diffusion and virality of information and (virtual) goods. In fact, virtual worlds have been used in the past for studying the actual virality of diseases (Boman and Johansson 2007; Lofgren and Fefferman 2007). Additionally, virtual worlds have been used as laboratories and experimental platforms for studying economic behaviour and decision-making, areas such as the spread of virtual goods within virtual worlds (Bakshy, Karrer, and Adamic 2009; Huffaker et al. 2011), labour markets (Horton, Rand, and Zeckhauser 2011), moral hazard (Tolvanen 2016), trust (Fiedler and Haruvy 2009; Fiedler, Haruvy, and Li 2011), avatar markets (Castronova 2001), and market efficiency (Golde 2008). A wide area of research based on a content within virtual worlds and online social networks is related to education (Baker, Wentz, and Woods 2009; Warburton 2009), collaborative learning (Lytras et al. 2015; Różewski et al. 2015) with special focus on dynamic experience management (Riedl et al. 2008), interactions (Petrakou 2010), engagement (Hamari et al. 2016), knowledge acquisition (Jankowski et al. 2015), multilayer knowledge diffusion (Rożewski and Jankowski 2015), social learning innovations (Lytras et al. 2014), web-based knowledge exchange (Filipowski et al. 2012), and online marketplace (D'Avanzo and Kuflik 2013).

During recent years, virtual goods have become a major category of consumption. The global value of the virtual goods market was at $14.8 billion in 2012 and was forecasted to continue rising in the near future (TechNavio 2013). Selling virtual goods has become a prominent business model for otherwise free online games and virtual worlds (Alha et al. 2014, 2016; Hamari and Järvinen 2011; Hamari and Lehdonvirta 2010). Thus, understating why people purchase and pass on virtual goods is a pertinent practical issue for virtual service operators (Hamari 2015). Virtual goods commonly refer to virtual objects, such as avatar clothing, weapons, virtual furniture, currencies, characters, and tokens. What sets virtual goods apart from digital goods, such as music and

CONTACT Jarosław Jankowski jjankowski@wi.zut.edu.pl

photos, is that they are rivalrous; virtual goods cannot be duplicated in the same sense as digital goods and they exist in the virtual world (Fairfield 2005; Hamari and Lehdonvirta 2010; Harviainen and Hamari 2015; Lehdonvirta 2009). As with any goods, an important aspect of virtual goods along with their functionality is their appearance. Especially in the content of this study, it is interesting how visual appearance may afford an aesthetic experience, experiences of provenance (Lehdonvirta 2009), and, therefore, increased diffusion.

While a plethora of research on motivations to purchase virtual goods exists (see e.g. Hamari and Keronen 2016), research on why users trade and distribute virtual goods among themselves is currently scarce. Even though both virality and virtual goods have become notable veins of research during the last decade, almost no research has been conducted on the merging of the two areas (Bakshy, Karrer, and Adamic 2009; Huffaker et al. 2011). Moreover, a notable gap in virality research exists concerning the role of the content characteristics, mechanics of diffusion, content visibility, and presentation layer. Earlier studies focused on the structure of social networks (Bampo et al. 2008), homophily (Aral, Muchnik, and Sundararajan 2009), or emotions (Stieglitz and Dang-Xuan 2013). Research related directly to diffusion of content in virtual worlds focused mainly on social aspects (Bakshy, Karrer, and Adamic 2009; Huffaker et al. 2011) or plagues and their similarities to real diseases (Boman and Johansson 2007; Kafai and Fefferman 2010; Kafai, Quintero, and Feldon 2010; Neulight 2005). Therefore, the present study attempts to address these questions regarding overlapping research areas of virality and virtual economies. However, the research also separately contributes to both domains of virtual goods and content diffusion individually by addressing looming research questions in both. The present study more specifically compares processes of digital content diffusion with several possible scenarios. The diffusion processes of virtual goods within a virtual world and factors affecting its performance in terms of coverage, adoption, and engagement were analysed in detail.

We investigate the role of content characteristics, mechanics of transmission, incentives, multilayer influence, sender and receiver characteristics, the relationship between sender and receiver, and visibility of the content. A tracking system allowed for the monitoring of the diffusion process and gathering of information if the content was visible or not at the moment when information about it was sent between a sender and a receiver. It is equivalent to a real-world situation in which it is possible to talk with other people and share knowledge about new products without showing the product itself or when the product is presented to potential customer. The role of content visibility in the moment of infections on further product adoptions would be difficult to observe in the real world; however, virtual worlds deliver the ability to create laboratories to observe such behaviours.

This paper is organised as follows: the next section includes a literature review related to the diffusion of information and the measurement of the effectiveness of viral campaigns; Section 3 describes the experimental set-up and general results of the study; Section 4 discusses factors affecting the dynamics of content diffusion; Section 5 explains the differences between processes based on visible versus non-visible content; and Section 6 presents a summary of the study.

## 2. Related work

Information and content flow within online social networks are one of the most interesting topics in recent years in the area of social media analysis. Spreading information or ideas with the use of viral mechanisms and the propagation of the marketing content are these days 'outsourced' to the consumers (Phelps et al. 2004). Close interpersonal relationships and social influence help in the distribution of content since individuals are more likely to spread it within their social networks (Chiu et al. 2007; Harvey, Stewart, and Ewing 2011). Content distribution in such a way allows for the limitation of advertising avoidance, which is characteristic of traditional mass marketing methods (Wilbur 2008). Diffusion processes exploit homophily effects within social networks. This is supported by the fact that individuals tend to have close relations with those who share similar interests (McPherson, Smith-Lovin, and Cook 2001). While the advantages of strategies based on spreading content using social influence are widely discussed, factors affecting successful campaigns are difficult to generalise. The low possibility that successful campaigns can be repeated makes this strategy to be often treated more as art than science (De Bruyn and Lilien 2008). As a result, most campaigns fail to achieve the widespread coverage that was expected during their planning (Watts, Peretti, and Frumin 2007).

While it is very difficult to launch a successful campaign, research behind diffusion processes within social networks is very wide and interdisciplinary. Studies in this area are discussing, among other things, strategies for the selection of initial nodes to maximise the coverage of campaigns as seeding strategies (Hinz et al. 2011; Liu Thompkins 2012), the role of the content and network structures (Bampo et al. 2008; Liu-Thompkins 2012), factors motivating users to forward content (Ho and Dempsey 2010), the role of emotions (Dobele et al.

2007; Stieglitz and Dang-Xuan 2013), and other factors (Berger and Milkman 2012; Camarero and San José 2011). Together the seed selection area of influence maximisation is explored (Chen, Wang, and Yang 2009). The theoretical background of information diffusion processes is derived from earlier epidemic modelling (Sohn, Gardner, and Weaver 2013) or branching processes (Van der Lans et al. 2010). However, recently other models and approaches have been proposed, including a linear threshold model (Pathak, Banerjee, and Srivastava 2010), an independent cascade model (Kempe, Kleinberg, and Tardos 2003; Wang, Chen, and Wang 2012), and a q-voter model (Even-Dar and Shapira 2007). Research related to viral marketing and information diffusion is based on mathematical models with the use of agent-based simulations (Perez and Dragicevic 2009), field experiments (Touibia, Stephen, and Freud 2011), datasets from social networking platforms such as Twitter (Taxidou and Fischer 2014) and Facebook (Li et al. 2013), virtual worlds (Bakshy, Karrer, and Adamic 2009; Huffaker et al. 2011), and e-commerce systems (Leskovec, Adamic, and Huberman 2007). Recent research opens new directions towards temporal networks (Jankowski, Michalski, and Kazienko 2013; Michalski et al. 2014), multilayer networks (Salehi et al. 2015), adaptive approaches (Seeman and Singer 2013), targeted viral marketing (Mochalova and Nanopoulos 2014), and evolving strategies (Stonedahl, Rand, and Wilensky 2010).

Various content and information can be the subject of diffusion within social networks, including images (Totti et al. 2014), video (Boynton 2009; Nelson-Field, Riebe, and Newstead 2013), news (Yang and Counts 2010), rumours (Jin et al. 2013), information about promotions and offers (Leskovec, Adamic, and Huberman 2007), virtual goods in virtual worlds, and special effects that can be applied to avatars, hair styles, or clothing (Bakshy, Karrer, and Adamic 2009; Huffaker et al. 2011).

Diffusion of virtual content is also a pertinent area of interest for companies attempting to capitalise on the value of virtual goods. Indeed, virtual goods today are a fundamental revenue source for many online ventures and platforms, and perhaps therefore, also a clear increase in studies investigating purchases of virtual goods during the last decade has been observed (see
e.g. Hamari and Keronen 2016). The questions of why people purchase virtual goods has been thus far investigated from multiple perspectives, such as the attributes that make virtual goods appealing (Lehdonvirta 2009), game design that creates demand for virtual goods (Hamari 2011; Hamari and Järvinen 2011; Hamari and Lehdonvirta 2010), cultural and demographic aspects (Lee and Wohn 2012; Wohn 2014), experiences derived from the environment where the virtual goods are being used in (Alha et al. 2014; Animesh et al. 2011; Cheon 2013; Chou and Kimsuwan 2013; Cleghorn and Griffiths 2015; Guo and Barnes 2011, 2012; Hamari 2015; Han
and Windsor 2013; Huang 2012; Kim 2012; Lee and Wohn 2012; Lin and Sun 2011; Liu and Shiue 2014; Luo et al. 2011; Paavilainen et al. 2013; Park and Lee 2011), customer lifetime value (Hanner and Zarnekow 2015), technology acceptance and planned behaviour (Cha 2011; Domina, Lee, and MacGillivray 2012; Gao 2014; Hamari and Keronen 2016; Mäntymäki and Salo 2011, 2013; Kaburuan, Chen, and Jeng 2009; Wang and Chang 2014).

Virtual goods are implemented on social platforms, online games, and virtual worlds, and built-in mechanics make it often possible to transmit virtual goods between user accounts. Even though information diffusion is one of the key topics of other research related to social media, a very limited number of efforts have attempted to investigate the diffusion of virtual goods among users of a virtual world (Bakshy, Karrer, and Adamic 2009; Huffaker et al. 2011). Environments of this type deliver the opportunity to study viral marketing processes because all interactions between users are potentially available for collecting, tracking, and analysing, and as a result, each stage of a diffusion process can be measured and evaluated. Earlier research showed that the adoption rate of content within virtual worlds, such as Second Life, increases together with the number of adopted friends from social networks (Bakshy, Karrer, and Adamic 2009). Other findings from the same study are related to the higher dynamics of content sharing among friends, however with a more limited audience and a lower reach in such a diffusion process. The authors emphasise that within the virtual world, it is possible to track transfers of content with detailed user behaviours focusing on not only the reach and the number of adopters (Bakshy, Karrer, and Adamic 2009). Another Second Life study related to virtual goods adoption performed by Huffaker et al. (2011) was based on the diffusion of digital assets among members of a group. It focused on the role of homophily in the adoption of ideas and innovations that followed (Jackson 2010). Social influence and social ties on the spread of information were analysed as well with the theoretical background derived from the diffusion of innovations (Rogers 2010). Two types of virtual goods were analysed, including internal bookmarks in the form of landmarks and gestures responsible for the sequences of movements or multimedia effects. Research showed the high importance of the crowding factor that represented the percentage of adopters in each group and group similarities, while the number of adopted friends had a lower impact on adoption. Apart from the crowding factor, the analysis covered item popularity, early and late adoptions, and the role of group characteristics.

Another aspect related to virtual worlds and content diffusion is based on virtual plagues affecting the players in massively multiplayer online games or virtual worlds (Boman and Johansson 2007). Specific extensions and modules with the ability to spread content between users were implemented for various purposes in games such as The Sims or World of Warcraft. They were transmitted among players even beyond the control of the platform operator and beyond the initial scope and functionality, for example guinea pig mod in The Sims or Corrupted Blood in the World of Warcraft. Boman and Johansson (2007) emphasise that virtual plagues can be implemented for more interesting game play and to increase player's engagement within the system. The authors discussed the advantages of micro-modelling of the virtual plagues over macro-modelling approaches for real diseases because of the low importance of virtual or real location, heterogeneous nature of players, lower risk avoidance in games than in real life, and multilayer or multi-channel spread of virtual diseases.

Lofgren and Fefferman (2007) discussed epidemics within the World of Warcraft from the point of view of simulation studies close to real-world diseases. The authors argue that even though observed epidemics had very high reproductive rates and dynamics, it would be possible to adjust transmission rates closer to real-world parameters. Virtual epidemics could be used in a simulation environment to test various scenarios and behaviours with different probabilities of transmission. The authors emphasise that typical simulation models lack the variability and unexpected outcomes not related directly to the disease, but to the nature of the agents it infects and the connections between them. Simulations based on virtual worlds and games are more close to reality. Specifics of virtual worlds make it possible to treat them as a learning environment for the discovery of the mechanics behind the spread of infectious diseases. Virtual plagues have similar perceived properties to real diseases (Neulight 2005). A study performed by Kafai and Fefferman (2010) was based on the Whyville virtual world and the Whypox virtual epidemic spread among avatars, which was further extended by Kafai, Quintero, and Feldon (2010). The authors showed the potential of virtual environments as learning laboratories for acquiring knowledge about infectious diseases and their mechanics. Data from virtual worlds make it possible to use micro-modelling (Boman and Johansson 2007). Differences between real and virtual diseases are difficult to model with the use of typical epidemiological models such as susceptible-infected-recovered (SIR) (Anderson and May 1991).

# 3. Experimental results and analysis of diffusion processes

## 3.1. The characteristics of an analysed virtual world

Experiments were conducted using the multiplayer system Timik.pl based on the isometric graphical virtual world that is targeted at teenagers in Poland and through the use of specially designed mechanics of content propagation within the social network. At the time of the study, more than 850,000 unique accounts had been created in the system. The virtual world is based on a freemium/free-to-play business model (see, e.g. Alha et al. 2016; Anderson 2009; Hamari and Lehdonvirta 2010) that gives free access to the world along with basic functions, but users can select to purchase more advanced features using mobile phone-based micro-transactions. For example, all users have some ability to customise the appearance of their avatars, but those wishing to create an avatar with a more unique appearance could do so by purchasing premium avatar packs. Moreover, while all users are able to meet in public rooms, users with premium accounts are able to purchase and furnish private rooms where their avatars can congregate. Communication between users occurs in the form of chats that take place in virtual public and private rooms with built-in dedicated functionalities for shops, schools, or restaurants. Examples of this are shown in Figure 1.

## 3.2. Multilayer social networks within virtual world

Usually in real world people are connected with many types of relations such as friendship, family ties, classmates, co-workers, acquaintance, etc. The same goes for our interactions on the Internet, people can be friends on Facebook, follow each other on Twitter, subscribe to channels on YouTube, and so on. In order to fully utilise the multitude of different ties, multilayer networks have been introduced (Boccaletti et al. 2014; Bródka and Kazienko 2014; Bródka et al. 2012; Kivelä et al. 2014; Musial et al. 2014). In this type of networks, different relations are represented as layers. Each layer is in fact a simple social network, where nodes represent users and links represent one type of relation between users. Multilayer networks are still quite new concept undergoing a lot of research. One of the areas currently studied

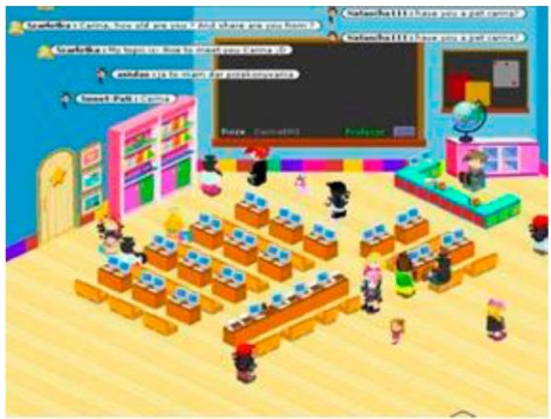
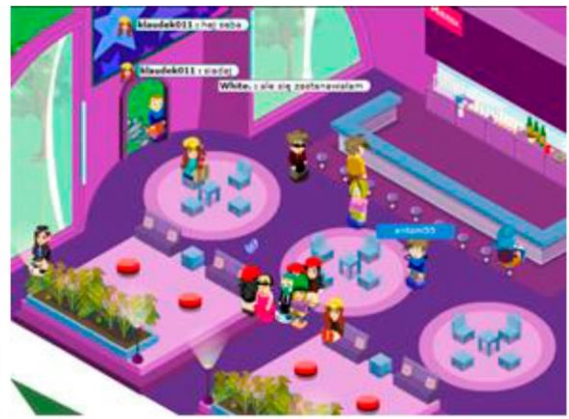

a) Virtual school and blackboard usage

b) Users meeting during a virtual party

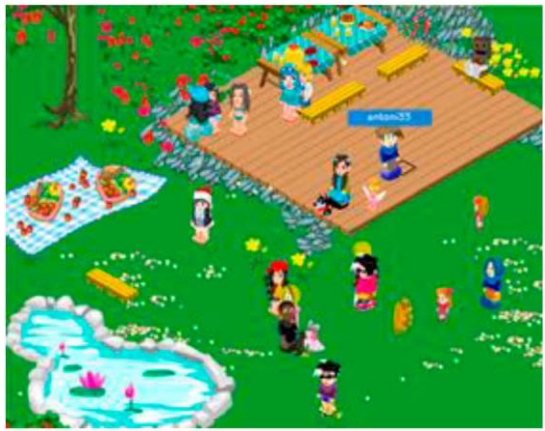
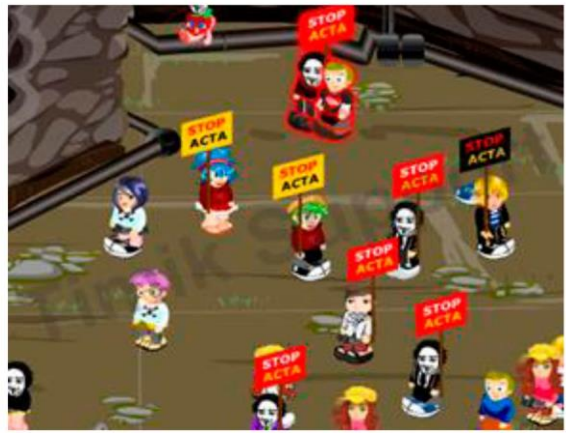

c) Virtual picnic and virtual goods usage

d) Protest with the use of distributed virally masks

Figure 1. Screenshots from examined virtual world and dedicated virtual rooms: (a) users attend lessons within the virtual school using the blackboard module with restricted access to experienced users; (b) virtual party in a dedicated public room with an available menu with virtual foods and drinks; (c) meeting of users at a public picnic presenting virtual goods from their inventories, such as animal avatars and hats; and (d) meeting of users with virally spread masks.

is spreading processes in multilayer networks (Salehi et al. 2015).

The analysed virtual world has several layers as well. These layers correspond to different types of interaction between users within the system. The first layer is based on connections between friends (friends lists), which is a result of invitations sent and received from other users. When a user accepts the invitation, he or she is added to the friends list of the inviting person and vice versa, the inviting person is added to the target user's list as well. The overall number of 12,603,439 friend connections (network edges) in the system databases is registered with 624,352 nodes representing connected users. However, connections in this layer do not represent close relations between users, because most users created these connections to get a high number of friends (for some users more than 5000) on the list, and acceptance rates for invitations were higher than 80%. The second network layer is based on messages sent between users through the use of the internal messaging system. Private messages are the alternative to communication using the open chat with messages visible to all users gathered in the room. The functionality of the messaging system is similar to email communication with the ability to send messages to both online and offline users. The only way to read messages is to log in to the system. It is possible to send messages only to users on the friends list. A total of 26,284,936 messages were sent between 364,791 unique users. The third layer within the system is based on the transactions and transfers of virtual currency between users. Virtual currency is used for purchasing avatars and other virtual goods such as food and paying for services in private restaurants, etc. There were a total number of 753,180 transactions between 112,112 unique users engaged in these transactions. The fourth layer is based on social activity within private virtual rooms organised by users as private houses, apartments, restaurants, schools, shops, etc. Rooms are a popular place for organising events and meetings. A total number of 16,464,000 visits to private rooms by 282,514 unique visitors were registered. Because of the limited role of the friends list, the evaluation of social relations for multilayer influence and diffusion was based on the analysis of messages, transactions, and social activity within private rooms.

### 3.3. Experiment design and set-up

Experiments were performed with the support of the designed content distribution system that had the ability to monitor the spread among users and identify factors affecting the dynamics of the process. The content subject to diffusion was in the form of the most popular virtual good available to users – special thematic avatars, making users more unique within the virtual world. Typical user avatars were available within free basic accounts. Their extended versions with higher quality were available for premium paid accounts. Apart from the two types of accounts, it was possible to buy avatars with the use of virtual currency in a dedicated shop. Avatars designed for diffusion experiments were unique and different from earlier available ones within the system. The only way to have them in inventory was receiving them from the other user. Experiments were conducted with the different diffusion mechanisms and content characteristics to observe how they affect the dynamics of diffusion and user behaviour. Two content diffusion mechanisms were implemented. First, treating as low resistance (LR) was an easy way for content transmission and users need to simply point and click on the recipient's avatar. Second, high-resistance (HR) transmission was based on social connections within special messages sent between users through the use of the internal messaging system. It was only possible for users to use this if they had each other on their friends list. Another analysed factor affecting the diffusion process was content visibility during transmission. This was based on a fact that users can wear various avatars from their inventory and even if they received an experimental avatar from other users, they do not necessarily have to wear it and present it to other users. Diffusion can be performed when the avatar is currently being worn and visible to the recipient; however, it can be transmitted when it is not worn and visible as well. Figure 2 illustrates the mechanics for LR (a) and HR content diffusion (b).

LR diffusion with content being redirected by clicking on a target avatar is illustrated in Figure 2(a). Diffusion starts from seed user (S) and can be performed with two types of transmissions. First, transmission $T^0$ takes place when the sender, in the moment of content transmission, uses (is wearing) another than the transmitted avatar; that is, the transmitted avatar is not visible to the receiver. Transmission $T^1$ takes place when the user is wearing the diffused avatar. The symbol $T^s$ denotes the diffusion with the state s={0,1} from user $i$ to user $j$. The symbol $U^i_j$ denotes user $j$ with content received from user $i$. After receiving content, the receiver has the ability to spread the content among friends in the same way – by clicking on their avatars with transmission $T^1$ or $T^0$. The received content can be adopted; that is, the user will wear the received avatar at least once that is denoted by $A(U_j)$. HR transmission is illustrated in Figure 2(b) and it requires more effort from the user to pass on the content. It is transmitted with the special code used within private messages. The process of sending content is based on three steps. In Step 1, the recipient of the message is selected from the friends list; in Step 2, a message is prepared and the activation code is inserted; and in Step 3, a message is sent to the recipient. Transmission with non-visible and visible content ($T^0$, $T^1$) takes place in the same way for LR content, and the receiver can choose to adopt it or not.

Another aspect taken into account stems from the motivation of users to spread the content. The first way to spread the content is based on a natural interest in the content self or the natural relation between users, without any other support. The only reason to redirect the content is a sender's intention to share the content with friends or to fulfil the requests from friends asking for new content. Apart from natural diffusion, the process can also be supported by incentives and users can receive rewards for spreading the content. Another factor was the difference in quality, with the similarity to basic and premium content. The last considered factor was emotional appeal of user-generated content. According to the above assumptions, five content elements $C_1$–$C_5$ in the form of user avatars with specifications as shown in Table 1 were implemented. The different approaches for analysing the mechanics and content characteristics were used.

Table 1. Characteristic of content used for diffusion experiments.

| Id | Content type | Incentivised | Quality | Mechanics |
|---|---|---|---|---|
| $C_1$ | Occasional avatar related to the special Halloween event | No | Premium | LR |
| $C_2$ | Occasional avatar related to the special Halloween event | Yes | Premium | LR |
| $C_3$ | Winter avatar for winter theme and seasonal room | No | Basic | HR |
| $C_4$ | Thematic avatar for special anniversary event | No | Premium | HR |
| $C_5$ | User-designed avatar with emotional appeal (Guy Fawkes mask) | No | Premium | HR |

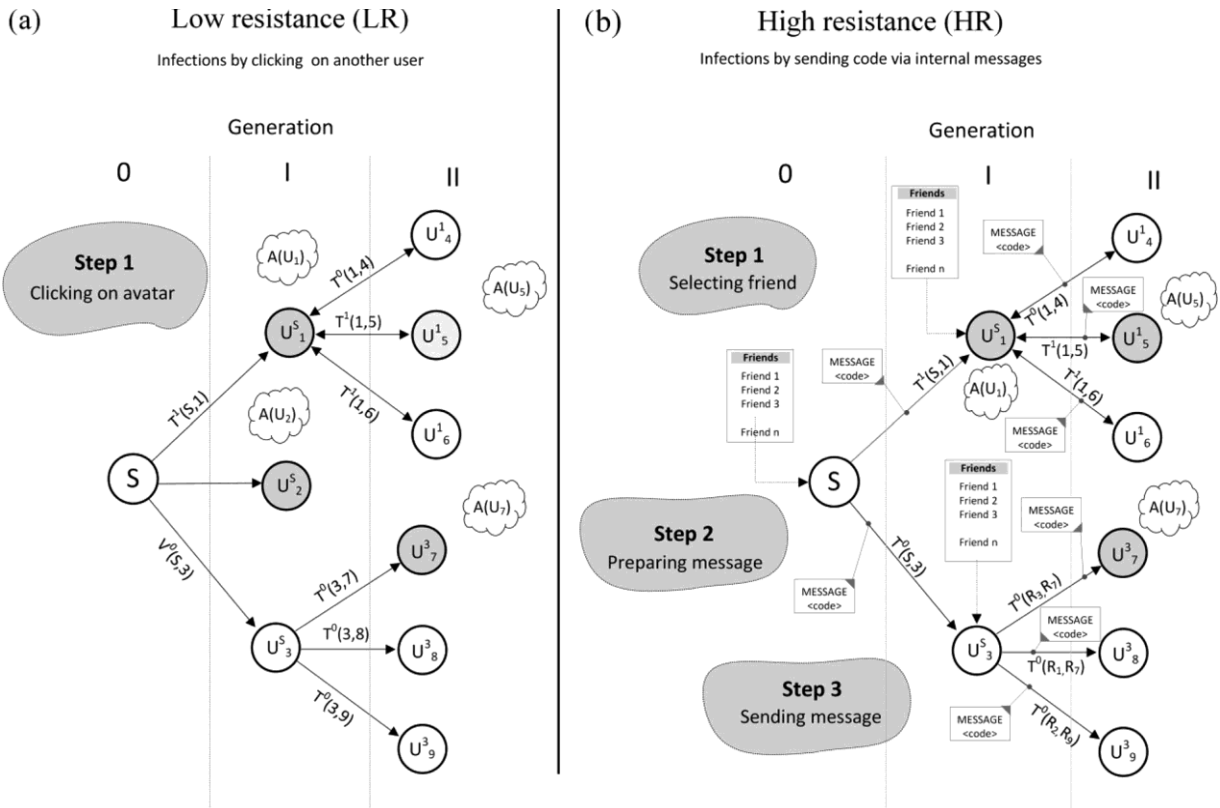

Figure 2. Differences between LR and HR transmission.

LR content transmission mechanics based on clicking on the recipient was used for content $C_1$ and $C_2$. The spread of content was limited to visible and logged in users who were occupying the same public or private room. Only 25 users can be in the same room at the same time. The HR diffusion of content $C_3$–$C_5$ was based on social connections and messages sent to friends. Content $C_1$ spreads without any additional support or motivation apart from natural user interest in new content. For content $C_2$, incentivised competition with prizes for users infecting the highest number of friends was used. Content $C_3$ was treated as a lower quality content, similar in quality to avatars available for free with basic accounts. In contrast, content $C_4$ was treated as similar in quality to the avatars available in premium accounts. Content $C_5$ with emotional appeal was designed by the users and was related to the special event organised from their initiative.

### 3.4. General analysis of diffusion processes

In the first stage of analysis regarding the spreading of content $C_1$–$C_5$, different mechanics in the time dimension for 60-second intervals was performed. Receiving content from another user was treated as an infection (I), further usage of content (wearing the avatar) was interpreted as an adoption (A), and sending the content to other users represented an engagement (E). The dynamics of the diffusion processes were characterised by three main factors: the cumulative number of infections (I), the number of adoptions (A), and the number of users engaged in redirecting the content (E). The results from the five processes for content $C_1$–$C_5$ are illustrated in Figure 3(a–e).

Different dynamics as well as the relationship between the number of infections and adoptions were dependent on the mechanics used and the type of content. An average reproductive ratio (ARR) represents the number of users to which infected and engaged users passed content. For LR content $C_1$, the $ARR_1$ reached an average of 3.66 infections per user. The reproductive ratio for incentivised content $ARR_2$ reached an average of 6.52 infections sent per user. The same factors were analysed for the content with the HR mechanics $C_3$–$C_5$. For lower quality content $C_3$, the total $ARR_3$ was equal to 2.91; for content $C_4$, the reproductive ratio $ARR_4$ was 2.93; and for content $C_5$, the reproduction rate reached 2.84.

For a more detailed evaluation of the diffusion process, three performance factors, as presented in the charts in Figure 3(f–h), were used. The adoption rate $AR_i$ represents the relationship between the number of adoptions and the number of content infections $i$. The relationship between the number of engaged users and the number of adoptions for each content $i$ is represented by the engagement rate $ER_i$. The relationship between the number of adopted users and the number of engaged users for each content $i$ is represented by the conversion rate $CR_i$, showing the conversion of users from engaged to adopted. For LR content that is easy to spread, the relationship between adoptions and the number of infections reached an average adoption rate of $AR_1 = 17.05\%$. The engagement rate reached $ER_1 = 27.33\%$ and the observed conversion rate $CR_1$ was equal to 62.37%.

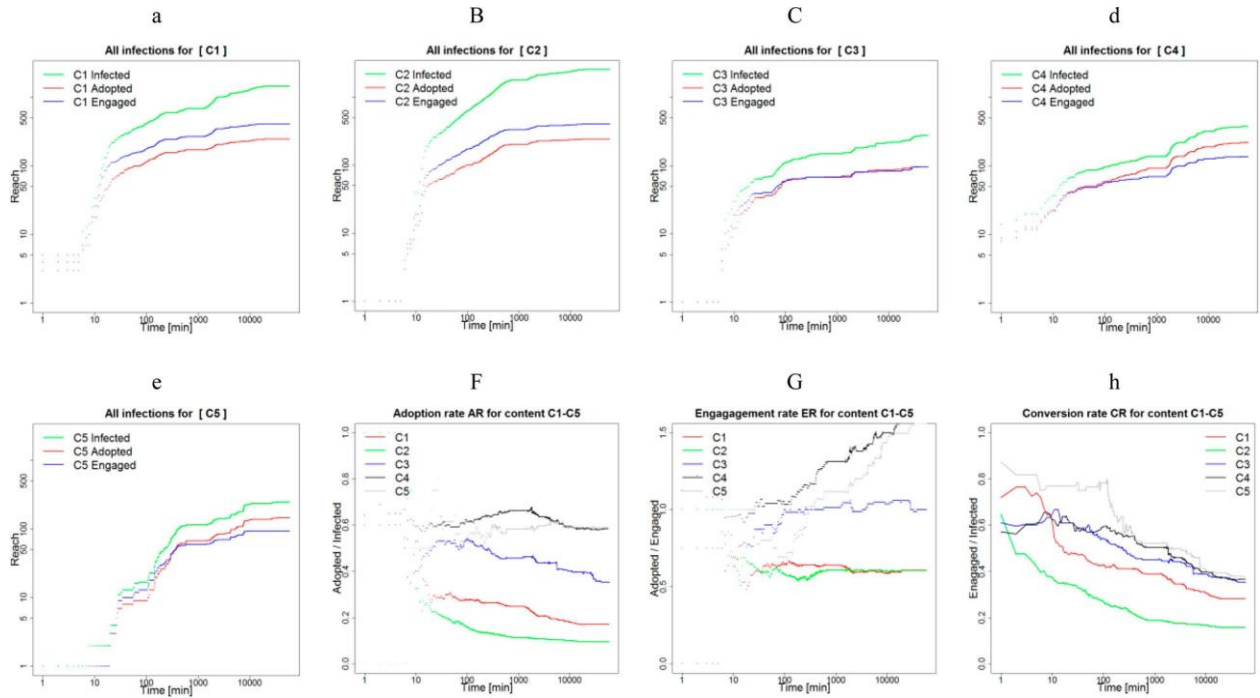

Figure 3. (a–e) Dynamics of the diffusion processes for content $C_1$–$C_5$ based on the incremental number of infections, adoptions, and engaged users; (f) AR is showing the relation between the number of adopted users and the number of infections for each content $C_1$–$C_5$; (g) ER based on the relationship between the number of engaged users and the number of infections for each content $C_i$; and (h) CR is the relationship between the number of adopted users and the number of engaged users for each content $C_1$–$C_5$.

Incentivised LR content $C_2$ achieved a 9.5% adoption rate, and the number of engaged users among all those infected was equal to 15.34%, with a 61.95% conversion rate for $CR_2$. HR content with low quality reached an adoption rate of 35%, and 34.33% of infected users engaged in content spreading. The conversion rate reached 101.95%, with the number of adopters higher than the number of engaged users. Higher quality content, $C_4$, was adopted by 57.80% of users among all those infected, and 34.17% of users engaged in content spreading. The conversion rate $CR_4$ reached 169.13%.

Table 2. Structure of generations of infections for content $C_1$–$C_5$.

| Generation | Content $C_1$ | | | Content $C_2$ | | | Content $C_3$ | | | Content $C_4$ | | | Content $C_5$ | | |
| --- | --- | --- | --- | --- | --- | --- | --- | --- | --- | --- | --- | --- | --- | --- | --- |
| | I | A | E | I | A | E | I | A | E | I | A | E | I | A | E |
| 0 | 24 | 9 | 4 | 39 | 11 | 2 | 5 | 1 | 1 | 1 | 17 | 1 | 4 | 3 | 1 |
| 1 | 102 | 35 | 13 | 187 | 30 | 8 | 33 | 16 | 4 | 1 | 25 | 1 | 5 | 4 | 2 |
| 2 | 284 | 65 | 54 | 552 | 64 | 49 | 68 | 22 | 20 | 3 | 38 | 1 | 8 | 8 | 5 |
| 3 | 394 | 58 | 98 | 782 | 55 | 110 | 74 | 33 | 26 | 23 | 35 | 2 | 25 | 16 | 7 |
| 4 | 266 | 35 | 89 | 450 | 37 | 102 | 70 | 21 | 26 | 52 | 42 | 10 | 36 | 19 | 11 |
| 5 | 179 | 23 | 59 | 251 | 19 | 54 | 38 | 8 | 17 | 53 | 29 | 24 | 38 | 23 | 14 |
| 6 | 86 | 11 | 36 | 137 | 14 | 32 | 11 | 4 | 8 | 65 | 35 | 22 | 41 | 29 | 17 |
| 7 | 50 | 9 | 19 | 47 | 4 | 18 | 1 | 0 | 1 | 70 | 21 | 22 | 43 | 31 | 15 |
| 8 | 48 | 2 | 14 | 27 | 3 | 5 | – | – | – | 42 | 8 | 21 | 44 | 11 | 12 |
| 9 | 13 | 0 | 8 | 51 | 4 | 4 | – | – | – | 64 | 2 | 18 | 14 | 9 | 9 |
| 10 | 3 | 0 | 2 | 6 | 0 | 4 | – | – | – | 49 | 0 | 19 | 12 | 2 | 2 |
| 11 | – | – | – | 7 | 0 | 1 | – | – | – | 10 | 0 | 5 | – | – | – |
| 12 | – | – | – | – | – | – | – | – | – | 3 | 0 | 3 | – | – | – |
| Total | 1449 | 247 | 396 | 2536 | 241 | 389 | 300 | 105 | 103 | 436 | 252 | 149 | 270 | 155 | 95 |
| Avg | 142.5 | 23.8 | 39.2 | 227.0 | 20.9 | 35.1 | 42.1 | 14.8 | 14.5 | 36.2 | 19.5 | 12.3 | 22.1 | 12.6 | 7.8 |
| SD | 132.6 | 23.8 | 34.3 | 259.6 | 22.7 | 39.5 | 29.5 | 11.5 | 10.2 | 26.5 | 16.6 | 9.4 | 17.4 | 10.9 | 6.0 |

User-generated content $C_5$ reached a 57.41% adoption rate $AR_5$ and 35.19% of infected users engaged in content spreading, with a 163.16% conversion rate $CR_5$.

## 3.5. Contagion waves and generation-based analysis

Analysis of the number of infections over time increases knowledge about dynamics; however, it does not deliver information about the structure of infections. The diffusion of five content types was analysed using an approach based on generations and parameters describing their characteristics derived from the branching processes (Van der Lans et al. 2010). A generation represents the number of transmissions required to reach a member along a chain of communications initiated by a single seed. An approach based on generations can capture structures of transmissions, which is not possible with a cumulative analysis based only on infections over time. In earlier research, epidemic theory based on branching models was used for modelling the characteristics of viral marketing campaigns (Stewart, Ewing, and Mather 2009). Table 2 presents the number of infections within the structure of generations for content $C_1$–$C_5$ with identified generation $G = 0$ as the first level of infections based on seeds. Further generations are created through the natural diffusion processes. The highest number of generations (12) was observed for content $C_4$ with high-quality content targeted to premium accounts. The smallest number of generations (7) was observed for lower quality content $C_3$ targeted to basic accounts.

For content $C_1$–$C_3$ the highest number of infections was observed in generation $G_3$ with 27.19%, 30.92%, and 24.67% of infections achieved in the third generation, respectively. For $C_4$ the highest number of infections was observed in the generation $G_7$ with 70 infections equal to 16.55% of all infections based on $C_4$ content. Content $C_5$ achieved the highest number of infections in generation $G_8$ (44) and that was 16.30% of the total number of infections for $C_5$ content. Further analysis includes parameters derived from epidemic modelling based on contagion parameter, epidemic intensity, and epidemic threshold (Stewart, Ewing, and Mather 2009). Contagion parameter denoted as $p$ describes the probability of transferring a viral message by an infected user. Epidemic intensity $\lambda$ represents the number of users reached. Epidemic threshold parameter (ETP) defined as $p * \lambda$ describes the progression of epidemics. Becker (1989) defined the relationship between the characteristics of a campaign and an ETP as subcritical (ETP < 1), supercritical (ETP > 1), and critical (ETP =1). Stewart, Ewing, and Mather (2009) presented a conceptual framework for viral marketing with the use of branching processes. The mathematical model presented by the authors for viral campaigns modelling was based on the deterministic model discussed by Frauenthal (2012), used by Anderson, May, and Anderson (1992) for modelling infectious diseases spreading, and extended later by Fulford, Forrester, and Jones (1997). Following earlier approaches, the analysis of viral processes was performed. ETPs, as well as $p$ and $\lambda$, were computed and are shown in Table 3 for all diffusion processes.

As Table 3 shows, the epidemic parameters change over time. In the analysis of $C_1$, the epidemic intensity $\lambda$ reached on average value of 3.47, and the ETP was supercritical in the first three generations, while the dynamics was dropped starting from generation $G_4$ with a slight increase to 0.96 and the supercritical level was almost reached in generation $G_{11}$. Content $C_2$ achieved better coverage in terms of the number of infections, but the structure of the generations was similar to that of the supercritical ETP in generations $G_0$–$G_3$, with dropping values starting from $G_4$ to an increase in ETP to supercritical levels in generations $G_9$ and $G_{11}$ and reaching values of 1.89 and 1.17, respectively. Content $C_3$ shows similar dynamics in terms of generations and supercritical values of ETP were observed until the third generation. Different characteristics were observed for content $C_4$ and $C_5$. Content $C_4$ achieved supercritical values of ETP in generations $G_0$–$G_7$. A higher number of generations with supercritical values >1 were observed for $C_5$ for generations $G_0$–$G_8$. Overall, the results based on this analysis of generations showed that incentives were not effective at increasing the number of generations characterised as supercritical according to the ETP for content $C_2$; however, the average for $\lambda$ was at the level of 7.67 and it was 2.21 times higher than that of $C_1$. The highest number of generations with supercritical ETP was achieved for high-quality premium content $C_4$ and user-generated content with more emotional appeal $C_5$.

## 4. The impact of user behaviour and content characteristics on the diffusion process

Available data from the system and the conducted experiments include characteristics of content, relations between the sender and the receiver, their behaviours, and activity within network layers with the variables in five groups $G_1$–$G_5$ presented in Table 4. The first group $G_1$ includes four variables related to a sender's characteristics. The input variable $U_5$ represents system usage since account creation until the time when an infection was sent, and it is based on the total number of logins to the system. Communication activity of the sender ($C_S$) represents the degree of the sender in the communication layer based on the usage of the internal messaging system for sending and receiving messages from other users. The model includes the degree within a social network based on the number of social connections represented by the number of friends acquired in the system ($F_S$), including inbound and outbound connections initiated by invitations sent between users. Social activity of the sender $A_S$ represents participation in meetings in private virtual rooms designed in the system by users. These include meetings in a sender's own room and attending meetings organised by other users in their rooms. In group $G_2$, the same parameters are gathered, and they represent content receivers' activities within the system. Group $G_3$ includes weighted representation regarding strengths of the relationship between senders and receivers based on their earlier communication prior to an infection. Weight $W_C$ shows the total number of messages sent between senders and receivers prior to an infection. Participation by content receivers in meetings organised by a sender and vice versa is represented by weight $W_A$. Transactional activity based on transfers of virtual currency between senders and receivers is represented by weight $W_T$.

Table 3. Characteristic of generations for content $C_1$–$C_5$ based on epidemic parameters.

| | $C_1$ | | | $C_2$ | | | $C_3$ | | | $C_4$ | | | $C_5$ | | |
|---|---|---|---|---|---|---|---|---|---|---|---|---|---|---|---|
| G | p | λ | ETP | p | λ | ETP | p | λ | ETP | p | λ | ETP | p | λ | ETP |
| 0 | 1.00 | 6.00 | 6.00 | 1.00 | 19.50 | 19.50 | 1.00 | 5.00 | 5.00 | 1.00 | 1.00 | 1.00 | 1.00 | 4.00 | 4.00 |
| 1 | 0.54 | 7.85 | 4.25 | 0.21 | 23.38 | 4.79 | 0.80 | 8.25 | 6.60 | 1.00 | 1.00 | 1.00 | 0.50 | 2.50 | 1.25 |
| 2 | 0.53 | 5.26 | 2.78 | 0.26 | 11.27 | 2.95 | 0.61 | 3.40 | 2.06 | 1.00 | 3.00 | 3.00 | 1.00 | 1.60 | 1.60 |
| 3 | 0.35 | 4.02 | 1.39 | 0.20 | 7.11 | 1.42 | 0.38 | 2.85 | 1.09 | 0.67 | 11.50 | 7.67 | 0.88 | 3.57 | 3.13 |
| 4 | 0.23 | 2.99 | 0.68 | 0.13 | 4.41 | 0.58 | 0.35 | 2.69 | 0.95 | 0.43 | 5.20 | 2.26 | 0.44 | 3.27 | 1.44 |
| 5 | 0.22 | 3.03 | 0.67 | 0.12 | 4.65 | 0.56 | 0.24 | 2.24 | 0.54 | 0.46 | 2.21 | 1.02 | 0.39 | 2.71 | 1.06 |
| 6 | 0.20 | 2.39 | 0.48 | 0.13 | 4.28 | 0.55 | 0.21 | 1.38 | 0.29 | 0.42 | 2.95 | 1.23 | 0.45 | 2.41 | 1.08 |
| 7 | 0.22 | 2.63 | 0.58 | 0.13 | 2.61 | 0.34 | 0.09 | 1.00 | 0.00 | 0.34 | 3.18 | 1.08 | 0.37 | 2.87 | 1.05 |
| 8 | 0.28 | 3.43 | 0.96 | 0.11 | 5.40 | 0.57 | – | – | – | 0.30 | 2.00 | 0.60 | 0.28 | 3.67 | 1.02 |
| 9 | 0.17 | 1.63 | 0.27 | 0.15 | 12.75 | 1.89 | – | – | – | 0.43 | 3.56 | 1.52 | 0.20 | 1.56 | 0.32 |
| 10 | 0.15 | 1.50 | 0.23 | 0.08 | 1.50 | 0.12 | – | – | – | 0.30 | 2.58 | 0.77 | 0.14 | 6.00 | 0.86 |
| 11 | – | – | – | 0.17 | 7.00 | 1.17 | – | – | – | 0.10 | 2.00 | 0.20 | – | – | – |
| 12 | – | – | – | – | – | – | – | – | – | 0.30 | 1.00 | 0.30 | – | – | – |
| Avg | 0.29 | 3.47 | 1.23 | 0.15 | 7.67 | 1.36 | 0.38 | 3.11 | 1.65 | 0.48 | 3.35 | 1.72 | 0.46 | 3.02 | 1.28 |
| SD | 0.14 | 1.89 | 1.30 | 0.05 | 6.21 | 1.41 | 0.24 | 2.41 | 2.28 | 0.28 | 2.81 | 2.03 | 0.27 | 1.27 | 0.73 |

Another included factor related to the process of sending content is based on the visual layer and the visibility of content during infection, and it is represented by the $V$ variable with a value of 0 or 1. All users having experimental content had an ability to distribute content to friends with the ability to be in two discrete states.

State $V=0$ represents a situation when content is not presented to the receiver, but it is available in the sender's repository. It stands for a real-life situation when two persons share information about a product; however, the product is not showed by the owner to the potential candidate for infection. State $V=1$ represents a situation when the content is visible to a potential candidate before an infection. Group $G_5$ represents variables with the potential to relate the social influences to the role of adopted friends in the process within three layers of social connections available in the system. These are based on the communication activity $L_C$, social activity $L_S$, and transactional activity $L_T$.

### 4.1. Factors affecting adoptions for each campaign

In the first stage, the analysis of the processes of spreading content was performed to identify factors affecting the adoption of receivers after infection resulting in content usage. Due to the different mechanics used and the type of distributed objects, separate regression models were built for each content $C_1$–$C_5$ based on variables from Table 4. Results in terms of statistical significance and the role of each input variable played are presented in Table 5. For content $C_1$ system usage $U_S$ by the sender affected the adoption ($p = 0.04$) the same as the social activity within the system $A_S$ with $p < 0.01$. These results are an indication of the social influence on the sender from experienced users within the system with high number of logins and social activity. On the receiver side ($G_2$), social activity positively affected content adoption ($p < 0.01$) at the same level as content visibility $V$ ($p < 0.01$). Weights of the relationship between senders and receivers did not affect the results. Adoption among friends in communication and social layers positively affected the results of the infection. The number of friends in the communication layer $L_C$ positively affected the adoption rate of the receiver ($p < 0.01$) in the same way as number of friends in the social activity layer $L_S$ did ($p < 0.01$). For content $C_2$ with incentivised diffusion process, characteristics of sender $G_1$ did not affect the adoption of the content. The degree in social connections layer of the receiver $F_R$ affected it ($p < 0.01$) as well as the visibility of the content ($p < 0.01$). For content $C_2$, the role of the weights on the relation between the sender and the receiver was observed with statistical significance for communication ($p < 0.02$) and social activity ($p < 0.02$). The number of adopted friends in both the communication and social activity layers affected the results in terms of content adoption by a sender. The distribution and adoption of low-quality content $C_3$ were related to the sender's social activity ($p < 0.04$), and characteristics of the receiver did not affect the results. The significance of content visibility was identified with $p < 0.01$ as well as significance communication ($p < 0.03$) and social activity layers ($p < 0.01$). For content $C_4$ and $C_5$ the significant role of the visual layer was identified ($p < 0.01$ and $p < 0.01$). Only one factor affected the results for $C_4$ and it was the sender's degree in the friends' connections layer ($p < 0.05$). For $C_5$ the only one significant factor was the number of adopted friends in the communication layer ($p < 0.03$).

Table 4. Variables used in the model to detect factors significant for adoption and engagement.

| Id Group | Symbol | Description |
|---|---|---|
| $G_1$ Sender | $U_S$ | System usage by a sender based on the number of logins to the system |
| | $C_S$ | Communication activity of a sender based on the number of internal messages |
| | $F_S$ | Number of friends acquired by a sender prior to the infection |
| | $A_S$ | Social activity of a sender based on participation in meetings in virtual rooms |
| $G_2$ Receiver | $U_R$ | System usage by a receiver based on the number of logins to the system |
| | $C_R$ | Communication activity of a receiver based on the number of internal messages |
| | $F_R$ | Number of friends acquired by a receiver prior to the infection |
| | $A_R$ | Social activity of a receiver based on participation in meetings in virtual rooms |
| $G_3$ Weights | $W_C$ | Weight of sender–receiver relation based on communication and number of messages |
| | $W_A$ | Weight of sender–receiver relation based on social activity and meetings in private rooms |
| | $W_T$ | Weight of sender–receiver relation based on virtual currency transactions |
| $G_4$ Visibility | V | Content visibility during infection |
| $G_5$ Layers | $L_C$ | Number of adopted friends in the communication layer |
| | $L_S$ | Number of adopted friends in the social activity layer |
| | $L_T$ | Number of adopted friends in the transactional layer |

The main goal of this analysis was to detect if consistent factors affecting the adoption of all distributed content can be identified, regardless of what mechanics and type of content are used. Results show that characteristics of the sender in group $G_1$ and the receiver in group $G_2$ generally did not affect the results and content adoption by a receiver. The role of social activity of the sender represented by $A_S$ was identified as affecting the results only for $C_1$ and $C_3$. System usage of sender $U_S$ affected the adoption of content $C_1$ and the degree in social network $F_S$ positively affected the adoption by the receiver for $C_4$. Characteristics of receiver $G_2$ related to system usage $U_R$ and communication with other users $C_R$ did not affect any type of content. For receivers, only the number of friends $F_R$ affected the adoption of $C_2$ and social activity $A_R$ affected the adoption of $C_1$. The visibility of content $V$ affected the adoption of all types of content with both LR and HR mechanics. Weights of the relation between a sender and a receiver affected the adoption of content only for $C_2$, with statistical significance for communication $W_C$ and social activity $W_A$. Transactional activity between senders and receivers did not affect any of the content types. In terms of social connections in the three distinguished layers, the number of adopted friends in the communication layer $L_C$ was significant for four types of content $C_1$–$C_3$ and $C_5$. The number of adopted friends in the social activity layer LA related to meetings in private rooms positively affected the adoption of content $C_1$–$C_3$ and had no effect on the adoption of content $C_4$ and $C_5$.

## 4.2. Factors affecting engagement for each campaign

Regression models were used with the same set of parameters for the identification of factors affecting the engagement of receivers in sending content to other users. The results are presented in Table 6. The sender's usage of the system $U_S$ affected the engagement in the distribution of content by the receiver for $C_1$ and $C_3$ with values of $p = 0.01$ and $p < 0.01$, respectively. Communication activity with the use of internal messages was not significant for any of the content. The degree in friends layer $F_S$ affected only the diffusion of content $C_4$ with $p < 0.01$. Social activity affected content $C_1$ and $C_3$, with statistical significance at the levels of $p < 0.01$ and $p = 0.05$, respectively. Results for a sender's characteristics did not show a clear view of its role in the diffusion process. A receiver's characteristics in terms of prior to infection usage showed statistical significance for the distribution of three of the five content units $C_2$, $C_3$, and $C_5$, with $p$-values of <0.04, <0.01, and <0.02, respectively.

Communication activity $C_S$ with other users did not affect engagement in the distribution of any content. Social position and the degree in friends network $F_R$ were important for engagement in the distribution of content $C_1$ and $C_2$. The social activity $A_R$ of the receiver affected engagement in the distribution of content $C_1$. In terms of visibility of content during infection, it affected four out of the five diffusion processes ($C_1$, $C_2$, $C_4$, and $C_5$). The relationship between the sender and the receiver did not affect the engagement of the receiver in the diffusion process. The number of adopted friends in

Table 5. Results from the linear regression model for adoptions in terms of statistical significance.

| Group | Symbol | p-Value | | | | | F-test | | | | |
|---|---|---|---|---|---|---|---|---|---|---|---|
| | | $C_1$ | $C_2$ | $C_3$ | $C_4$ | $C_5$ | $C_1$ | $C_2$ | $C_3$ | $C_4$ | $C_5$ |
| Sender | $U_S$ | 0.04 | 0.10 | 0.25 | 0.31 | 0.34 | 4.06 | 2.67 | 1.31 | 1.03 | 0.91 |
| | $C_S$ | 0.64 | 0.14 | 0.40 | 0.06 | 0.47 | 0.21 | 2.18 | 0.70 | 3.65 | 0.53 |
| | $F_S$ | 0.71 | 0.56 | 0.50 | 0.05 | 0.14 | 0.13 | 0.33 | 0.46 | 3.97 | 2.23 |
| | $A_S$ | **0.01** | 0.82 | **0.04** | 0.13 | 0.87 | **17.01** | 0.05 | **4.39** | 2.28 | 0.03 |
| Receiver | $U_R$ | 0.30 | 0.52 | 0.83 | 0.65 | 0.54 | 1.09 | 0.42 | 0.05 | 0.20 | 0.38 |
| | $C_R$ | 0.15 | 0.92 | 0.43 | 0.79 | 0.22 | 2.05 | 0.01 | 0.63 | 0.07 | 1.49 |
| | $F_R$ | 0.10 | **0.01** | 0.42 | 0.56 | 0.29 | 2.73 | **32.40** | 0.51 | 0.34 | 0.49 |
| | $A_R$ | **0.01** | 0.86 | 0.29 | 0.08 | 0.90 | **13.75** | 0.03 | 1.11 | 3.16 | 0.02 |
| Visibility | V | **0.01** | **0.01** | **0.01** | **0.01** | **0.01** | **15.98** | **23.14** | **9.21** | **11.23** | **7.68** |
| Weight | $W_C$ | 0.57 | **0.02** | 0.16 | 0.88 | 0.70 | 0.33 | **5.46** | 1.96 | 0.02 | 0.15 |
| | $W_A$ | 0.73 | **0.01** | 0.70 | 0.80 | 0.06 | 0.12 | **5.95** | 0.15 | 0.07 | 3.54 |
| | $W_T$ | 0.23 | 0.23 | 0.53 | 0.24 | 0.47 | 1.47 | 1.42 | 0.39 | 1.38 | 0.53 |
| Layer | $L_C$ | **0.01** | **0.03** | **0.02** | 0.32 | **0.02** | **21.62** | **4.57** | **5.40** | 1.00 | **5.37** |
| Group | $L_S$ | **0.01** | **0.01** | **0.01** | 0.06 | 0.12 | **26.11** | **26.86** | **7.10** | 3.70 | 2.50 |
| | $L_T$ | 0.53 | 0.77 | 0.51 | 0.69 | 0.25 | 0.39 | 0.09 | 0.43 | 0.16 | 1.35 |

Note: Bold values are statistically significant with $p < 0.05$.

the layer based on the social activity ($L_S$) affected the engagement of the receiver of content $C_1$, while the communication layer $L_C$ and the social activity layer $L_S$ affected the engagement of diffusion of content $C_2$. For content $C_3$ and $C_4$, the number of adopted friends in network layers did not affect engagement in the sending of content to other users. For content $C_5$, the communication layer $L_C$ was identified as significant.

## 5. The comparative study of performance for visible and non-visible content

The performed analysis revealed that the fact whether the content is visible has an impact on both adoptions and engagement in digital content distribution in the virtual world. Further analysis is based on a comparative study of the spreading processes of content $C_1$–$C_5$, dependent on content visibility (Table 7).

The lowest share of spreading with visible content (avatar) during infection was observed for $C_1$ (8.97%). The highest proportion of visual infections was observed for $C_5$, with 57.04% of all infections performed with visible content. An average share of visible content for all five content units was at the level of 28.45%. The AR for visible content reached higher values than that for non-visible content for all types of content. For both visual and non-visual infections, the lowest adoption rate was observed for incentivised content $C_2$; that is, users were spreading content as much as they could in order to get points, but only 17% people were interested in that content. For each content, the AR with visible content was from 1.25 to 2.26 times higher than the AR for non-visible content. An average adoption rate for visible content (46.36%) was 1.51 times higher than that for non-visible content (30.64%). The engagement rate for visible content reached 38.07%, which was 1.48 times higher than that for non-visible content at 30.64% average ER. For visible content, the lowest ER was for incentivised content, with value 24.73% for visual infections and 13.76% for non-visual infections. Similarly to AR, the ER for visible content was always higher than that for non-visible content.

For each content, the relationship of the adoption rate between visible and non-visible content was analysed

Table 6. Results from the linear regression model for engagement in terms of statistical significance.

| Group | Symbol | p-Value | | | | | F-test | | | | |
|---|---|---|---|---|---|---|---|---|---|---|---|
| | | $C_1$ | $C_2$ | $C_3$ | $C_4$ | $C_5$ | $C_1$ | $C_2$ | $C_3$ | $C_4$ | $C_5$ |
| Sender | $U_S$ | **0.01** | 0.84 | **0.01** | 0.71 | 0.35 | **6.27** | 0.04 | **15.48** | 0.14 | 0.88 |
| | $C_S$ | 0.08 | 1.00 | 0.09 | 0.11 | 0.89 | 2.99 | 0.01 | 2.91 | 2.59 | 0.02 |
| | $F_S$ | 0.22 | 0.85 | 0.67 | **0.01** | 0.15 | 1.48 | 0.03 | 0.19 | **6.91** | 2.08 |
| | $A_S$ | **0.01** | 1.00 | 0.05 | 0.21 | 0.37 | **8.29** | 0.00 | 4.05 | 1.55 | 0.81 |
| Receiver | $U_R$ | 0.90 | **0.04** | **0.01** | 0.16 | **0.02** | 0.02 | **4.21** | **6.67** | 1.98 | **5.74** |
| | $C_R$ | 0.76 | 0.08 | 0.26 | 0.08 | 0.60 | 0.09 | 3.12 | 1.28 | 3.09 | 0.28 |
| | $F_R$ | **0.02** | **0.01** | 0.13 | 0.48 | 0.62 | **5.55** | **97.20** | 1.73 | 1.17 | 0.21 |
| | $A_R$ | **0.01** | 0.77 | 0.34 | 0.57 | 0.59 | **5.95** | 0.09 | 0.91 | 0.33 | 0.30 |
| Visibility | V | **0.01** | **0.01** | 0.25 | **0.02** | **0.04** | **9.41** | **12.18** | 1.33 | **5.86** | **4.30** |
| Weight | $W_C$ | 0.68 | 0.96 | 0.32 | 0.73 | 0.88 | 0.17 | 0.01 | 0.98 | 0.12 | 0.02 |
| | $W_A$ | 0.08 | 0.21 | 0.15 | 0.66 | 0.20 | 3.11 | 1.57 | 2.04 | 0.20 | 1.64 |
| | $W_T$ | 0.25 | 0.25 | 0.51 | 0.13 | 0.33 | 1.35 | 1.34 | 0.43 | 2.31 | 0.96 |
| Layer | $L_C$ | 0.20 | **0.03** | 0.09 | 0.33 | **0.02** | 1.63 | **4.92** | 2.92 | 0.96 | **5.44** |
| | $L_S$ | **0.01** | **0.01** | 0.39 | 0.89 | 0.90 | **7.23** | **27.54** | 0.75 | 0.02 | 0.02 |
| | $L_T$ | 0.86 | 0.20 | 0.11 | 0.31 | 0.73 | 0.03 | 1.61 | 2.57 | 1.02 | 0.12 |

Note: Bold values are statistically significant with $p < 0.05$.

Table 7. The performance in content diffusion for visible and non-visible content, with comparison of average reproductive rate (ARR), adoption rate (AR), engagement rate (ER), and conversion rate ($C_R$) between visible and non-visible content, and vice versa.

| Parameter | Visible content | | | | | Non-visible content | | | | |
|---|---|---|---|---|---|---|---|---|---|---|
| | $C_1$ | $C_2$ | $C_3$ | $C_4$ | $C_5$ | $C_1$ | $C_2$ | $C_3$ | $C_4$ | $C_5$ |
| Infections | 130 | 457 | 68 | 155 | 154 | 1319 | 2079 | 232 | 281 | 116 |
| Infections (%) | 8.97 | 18.02 | 22.67 | 35.55 | 57.04 | 91.03 | 81.98 | 77.33 | 64.45 | 42.96 |
| Senders | 51 | 44 | 25 | 69 | 59 | 367 | 369 | 91 | 107 | 58 |
| Adopted | 40 | 80 | 35 | 103 | 101 | 207 | 161 | 70 | 149 | 54 |
| Engaged | 53 | 113 | 29 | 63 | 64 | 354 | 286 | 73 | 85 | 31 |
| ARR | 2.55 | 10.39 | 2.72 | 2.25 | 2.61 | 3.59 | 5.63 | 2.55 | 2.63 | 2.00 |
| AR (%) | 30.77 | 17.51 | 51.47 | 66.45 | 65.58 | 15.69 | 7.74 | 30.17 | 53.02 | 46.55 |
| ER (%) | 40.77 | 24.73 | 42.65 | 40.65 | 41.56 | 26.84 | 13.76 | 31.47 | 30.25 | 26.72 |
| CR (%) | 75.47 | 70.80 | 120.69 | 163.49 | 157.81 | 58.47 | 56.29 | 95.89 | 175.29 | 174.19 |
| ARR comparison (%) | 70.92 | 184.35 | 106.69 | 85.54 | 130.51 | 141.00 | 54.25 | 93.73 | 116.91 | 76.62 |
| AR comparison (%) | 196.06 | 226.05 | 170.59 | 125.32 | 140.89 | 51.00 | 44.24 | 58.62 | 79.79 | 70.98 |
| ER comparison (%) | 151.91 | 179.74 | 135.54 | 134.37 | 155.51 | 65.83 | 55.64 | 73.78 | 74.42 | 64.30 |
| CR comparison (%) | 129.07 | 125.76 | 125.86 | 93.27 | 90.60 | 77.48 | 79.51 | 79.45 | 107.22 | 110.38 |

over time. The results are shown in Figure 2(a, b). The highest distance was observed for content $C_2$ and $C_5$. For content $C_4$, the adoption rate was similar for visible and non-visible content. Results show that visual factors positively affected all types of content and diffusion mechanics including LR and HR. While the AR shows a clear relationship to content visibility at the moment of infection, the impact of content visibility on the ER was at a lower level. The differences between visual and non-visual infections were observed for LR content $C_1$ and $C_2$. This is shown in Figure 4(f, g). For content $C_3$–$C_5$ with charts in Figure 4(h–j), the ER was similar for both visible and non-visible content.

### 5.1. Characteristics of senders and receivers

For senders with visible content, the following parameter values were calculated: platform usage, that is, the average number of logins, $U_S = 1139.03$; communication activity and the average number of messages, $C_S = 4966.49$; the average number transactions, $T_S = 154.76$; and social activity based on meetings in virtual rooms, $A_S = 2328.34$. These where compared to values calculated for senders with non-visible content $U_S = 1201.65$, $C_S = 5087.88$, $T_S = 164.13$, and $A_S = 2341.32$. Senders' parameters for visible content had slightly lower values. Higher differences were observed for content receivers. The values for receivers of visible content $U_S = 832.88$, $C_S = 3427.41$, $T_S = 105.11$, and $A_S = 1429.24$ were greater than those for receivers of non-visible content: 1.38, 1.36, 1.17, and 1.42 times, respectively. Experienced users noticed new content and showed interest in it, motivating content owners to share it with them. Lesser engaged users did not initiate the infections when the content was not visible. They were motivated by senders and social influence was observed from more experienced content senders to less experienced receivers. For both visible and non-visible content, senders had higher values of platform usage ($U_S$), communication activity ($C_S$), transactions ($T_S$), and social activity ($A_S$) than content receivers. For visible content, platform usage of the sender ($U_S$) was 1.61 times higher than that of the receiver's, communication activity ($C_S$) of the sender was 1.58 times higher than that of content receiver, transactional activity $T_S$ was 1.59 times higher, and social activity $A_S$ was 1.77 times higher. The same values analysed for non-visible content show that differences were higher. Platform usage of the sender was 2.22 times higher than that of the receiver's, communication activity of the sender was 2.2 times higher than that of the content receiver, transactional activity was 1.95 times higher, and social activity was 2.28 times higher. The analysis of adoptions for visual content shows that smaller differences between senders and adopters existed. Adoption was performed by users similar to senders. For visible content, the senders' parameters were 1.28, 1.34, 1.20, and 1.21 times higher than that of the adopted users for $U_S/U_R$, $C_S/C_R$, $T_S/T_R$, and $A_S/A_R$, respectively. While for non-visible content, differences between senders and adopters dropped to 1.71, 1.79, 1.46, and 1.71 for the same factors.

### 5.2. Relationships between senders and receivers

The relationships between senders and receivers were analysed in terms of sent messages, social activity, and transactions. The average number of messages between sender and receiver for infections with visible content was equal to 19.41 and for non-visible content it was equal to 25.55. This shows that when the content was not visible, word of mouth had an effect on infections, and sharing content with closer friends was observed. If the content was visible, infections were invoked between users with weaker relationships. Social activity between senders and receivers was higher for visible content with an average of 7.53 meetings in private rooms, while an average of 4.26 was observed for infections with non-visible content. The number of transactions between senders and receivers was similar for visual and non-visual infections with values of 1.07 and 0.93, respectively.

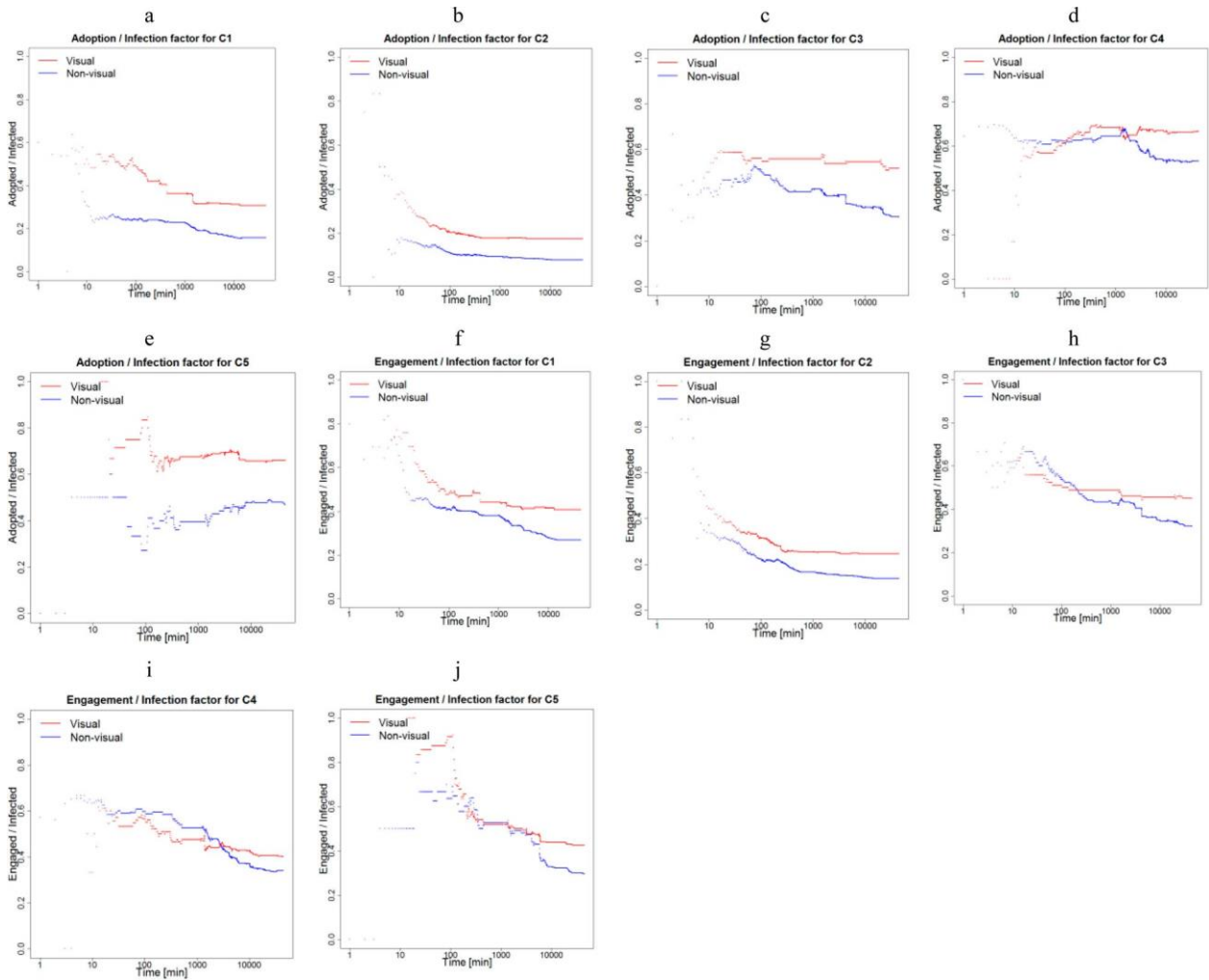

Figure 4. (a–e) AR for content $C_1$–$C_5$ with illustrated adoption rate for visible and non-visible content. (f–j) Engagement rate for content $C_1$–$C_5$ with separate charts for visual and non-visual infections.

### 5.3. Role of social influence in layers

The number of adopted friends in the social activity layer affected the adoption rate, that is, that user started using new content, for both visible and non-visible content. When content $C_1$ was visible during the infection, 28.16% of receivers adopted it despite the fact that they have not had any adopted friend in the social activity layer. When the number of adopted friends in the social activity layer was greater than 0, the AR increased 1.45 times to 40.74%. For content $C_2$, the AR increased from 13.61% to 47.17% (3.46×) if the receiver had adopted friends in the social activity layer. For content $C_3$–$C_5$ the adoption rate increased from 50.00%, 57.84%, and 63.25% to 58.33%, 83.02%, and 72.97%, respectively. When the content was not visible during the infection, adopted friends in the social activity layer had a higher impact on AR. For not visible content $C_1$–$C_5$, the adoption rate was 11.55%, 6.49%, 27.89%, 49.40%, and 44.00%, respectively, and it increased to 30.45% (2.64×), 17.90% (2.76×), 40.48% (1.45×), 58.41% (1.18×), and 51.22% (1.16×) when the receiver had at least one adopted friend in the social activity layer. The communication layer was less important for adoption than the social activity layer and the number of adopted friends on the communication layer did not affect the adoption rate. It is important to notice that on the social activity layer users are in the same private room and can see each other, while on the communication layer they extend messages and do not have to see each other; what is more, they do not have to be logged in to the system at the same time. Communication within the internal messaging system was massive and not necessarily related to close relations and social influence. For visible content, results showed the adoption rate to be 30.95%, 17.62%, 51.06%, 67.01%, and 67.31% for users not having any adopted friends in the communication layer. For users with at least one adopted friend, AR was equal to 25.00%, 25.00%, 52.38%, 65.52%, and 62.00%. No relation was observed for non-visible content as well. For users without adopted friends, AR was 17.35%, 7.98%, 31.74%, 51.30%, and 47.62%, while with at least one adopted friend AR was 6.47%, 4.41%, 26.15%, 55.12%, and 45.28%. For the transaction layer, the results were similar to that of the communication layer.

## 6. Discussion

The presented research has shown the abilities and directions for organising research within virtual worlds as digital experimental environments and as an extension to earlier studies (Neulight 2005). Capturing various aspects related to the transmission of information and diffusion processes together with all the interactions between users gave this study the unique ability to analyse in detail the spreading of content. Performed experiments have shown how micro-modelling and tracking of detailed events and interactions can be used for studying diffusion processes in terms of coverage, time, generations, and factors affecting performance. The used approach based on micro-modelling follows the study from Boman and Johansson (2007) related to the differences between real and virtual diseases and the difficulties with using typical epidemiological models such as SIR.

This study has delivered several extensions to earlier approaches. For example, weighted social connections based on the number of interactions between users were utilised, instead of the simple fact of connection between users like in most research (e.g. Huffaker et al. 2011). Huffaker et al. (2011), during content adoption analysis, focused on homophily and social influence, and not on the characteristics of the content and the mechanics of diffusion. In Bakshy, Karrer, and Adamic (2009), analysis of gestures in Second Life was performed, which did not take into account the fact that if and when the gesture was shown to the content receiver. This paper shows various strategies and ways of using a virtual world as a laboratory for conducting research in the areas related to virtual economies and the spreading of content with aspects difficult to measure in real-world systems.

### 6.1. The impact of content visibility

The only factor positively influencing adoption rate of all content types was their visibility prior to infection. In fact, the adoption rate for visible content was up to two times higher than that for non-visible content.

This exemplifies the metaphor related to the role of picture (typical representation of non-functional virtual goods) when compared to communication based on words (and actually on word of mouth typical for viral marketing) and textual messages are valid, not only in the real world but also in the virtual world. One of the reasons for the higher adoption rate of visible content was the pull effect (Król, Budka, and Musial 2014) and the initiative from potential receivers asking the owner to share the content. The visibility of an avatar had an additional informative function, showing users new possibilities within the system, and was an element of the knowledge-sharing process. Even for incentivised content, which had a low adoption rate for both visible and non-visible content during transmission, the adoption rate for visible content was two times higher than for non-visible content. Additionally, for non-visible content, word of mouth had an effect on infections and sharing content only with close friends, that is, users with whom the sender exchanged higher number of messages, while for visible content, it spread between users with weak relationships, that is, between users who rarely communicate with each other.

### 6.2. The impact of content type

While Bakshy, Karrer, and Adamic (2009) showed a limited audience and lower reach for the diffusion process among close friends, in this paper the adoptions and engagement rates were additionally analysed. Another aspect not analysed before is the mechanism of content distribution and how it affects to whom users send content. For example, HR content is spread among closer friends, while LR content is distributed randomly among greater numbers of 'unknown' users, which induces a much lower rate of adoption. Even though LR content (even without incentives) achieved nearly five times more infections than any HR content, the adoption rate for HR content was always at least two times higher. This is even more clear if we compare the HR content with the LR incentivised content in which the adoption rate was five times lower due to the fact that users were 'paid' for spreading the content and not for using it. Incentivised content achieved the highest number of infections, but the adoption rate was at the lowest level. The number of incentivised infections was nearly two times higher than that for other LR content (without incentives), but the total number of adoptions was smaller.

The best results in terms of the adoption rate were achieved for premium HR content, while LR mechanics lead to redirecting the content to users not interested in new content and can be treated by receivers as unsolicited messages and content. This is confirmed by the analysis of generations based on branching processes. Results showed that the number of generations with an ETP > 1 representing supercritical characteristics and highest dynamics was observed for premium content. Longer waves of infections, long-lasting interest in content, and the highest number of generations were observed for them. The epidemic intensity parameter was highest for incentivised content, but incentives did not positively affect the number of generations with high ETP.

### 6.3. The impact of network layer characteristics

Our research confirms that the adoption rate of content within virtual worlds increases together with the number of adopted friends from social networks (Bakshy, Karrer, and Adamic 2009). We presented an extension towards multilayer social influence analysis and the fact that connections among users in different layers have different impacts on influence. In the analysed system, social connections based on internal messages or transactions were less important for spreading than connections based on social activity in virtual rooms. The number of adopted friends, that is, friends who were wearing the content in the social activity layer, affected the adoption rate for all types of content. At the same time, the number of adopted friends in the communication layer affected the content adoption only in four content types. This may indicate that interactions on the social activity layer motivated users more than word-of-mouth communication, where a user has no ability to see content.

Results also showed that during campaigns two major dependencies for diffusion can be observed. The role that the visual layer had on accelerating the adoption rate is related to pull mechanics in content diffusion when the attention of the content receiver is increased (Król, Budka, and Musial 2014). When the content was visible, queries about the content were sent by experienced users to the adopters to share the content. In such a case, the interest of the potential receiver boosted the dynamics of the process. The confirmed role of the number of adopted friends in the network, especially on the communication and social activity layers, is very similar to the concept of diffusion models based on pull characteristics, for example, in the linear threshold in which the number of activated nodes influences non-activated nodes to adopt. Results show that both models and mechanics can be observed even within a single diffusion process. The role of the visual layer in adoption and engagement confirms the self-selection process, in which users potentially interested in the product are contacted by the product owner, and after infection, they are interested in both adoption and further distribution.

### 6.4. The impact of sender and receiver characteristics and the relation between them

Results showed that factors affecting the adoption of and engagement in content distribution were different for each content in terms of the sender's and the receiver's characteristics. System usage by a sender and social activity affected receiver adoption for LR content without incentives. The social activity of the sender affected the adoption of high-quality content with HR, while the number of friends played a role for content with lower quality. Receiver characteristics in terms of adoptions played a limited role for LR mechanics without incentives in relation to social activity and for incentivised diffusion in terms of number of friends. Relations between the sender and the receiver represented in the form of weights based on earlier activities affected the incentivised content in relation to communication and social activity. Social influence from the sender was limited and did not affect further adoptions. More important was the influence from friends on social networks, especially for private room visitors.

From the perspective of engagement in content distribution, the role of the sender was similar to that in adoption, and the same variables played significant roles for most content. For non-incentivised LR content, system usage and social activity of the sender were important; the same goes for low-quality content. Additionally, for low-quality content, the number of friends played an important role. The engagement rate of the receiver in the distribution of content was not affected by the characteristics of the sender for incentivised and user-generated content. The engagement rate was affected more by receiver characteristics than the adoption rate. The receiver system usage and his or her experience were affected by the engagement rate for most content units. Communication activity did not affect this behaviour, while the number of friends affected the number of infections for LR content. Social activity of the receiver affected content distribution for non-incentivised content with LR.

Surprisingly, the weight of the relationship based on prior communication between senders and receivers did not affect adoption. It was possible due to the massive usage of the messaging system among users with low social relations. Another possibility was the fact that communication between users takes place in the form of open chat, and it is the main medium for building social relations. The usage of messages from open chat was not useful for this study because they do not have recipients and are publicly visible to users. This did not affect diffusion or monitoring because transferring content was possible with the use of internal messages.

## 7. Summary

In this paper, we have discussed the role that the visibility of content and other factors have on the adoption of virtual goods. Other analysed factors include sender and receiver characteristics, the relationship between users, and the social influence on content diffusion (virility) in digital environments, or more specifically virtual worlds. The results show that prior visibility of content (equivalent to product presentations) before diffusion is the strongest predictor of content adoption and further diffusion across the networks. Among the 15 analysed variables which describe sender and receiver characteristics as well as the influence of different interactions between users (network layers), the only significant factor that always positively influenced the spread of content, adoption, and engagement was its visibility prior to infection. When a user can see the content, he or she is more likely to adopt the content

(start using it) and then engage in further distribution of that content. This mechanism can be further strengthened by social relations, that is, when users can see the content used by his or her friends. However, interestingly, a user is more likely to adopt content and engage in its distribution when he or she can see his or her friends from a social activity layer using it. However, adoptions among friends in communication and transaction layers do not increase adoption or engagement rates.

Besides visibility, few more factors have a limited influence on the spreading process. The first one is the ease with which users can transfer the content. LR content with the ability to be easily transmitted to all users, not only people from a friends list, reached higher coverage. However, without the influence from social connections, LR processes resulted in much lower adoption rates compared to HR content transmitted to friends only. The second one is that incentives increase the final reach of the spreading process, but adoption rates were much lower than that in an approach without incentives.

Future work will focus on community detection in each layer, inter- and intra-community processes, as well as the role of communities in each layer on the adoption dynamics.

## Funding

This work was partially supported by the European Union's Seventh Framework Programme for research, technological development, and demonstration [grant agreement number 316097] [ENGINE]; by The Polish National Science Centre [decision number DEC-2013/09/B/ST6/02317]; by the Faculty of Computer Science and Management, Wrocław University of Science and Technology statutory funds; and by the Finnish Funding Agency for Innovation [TEKES/40111/14] and [TEKES/40107/14].